\documentclass[preprint,12pt,3p]{elsarticle}

\usepackage{amssymb}
\usepackage{amsthm}
\usepackage{amsmath}
\usepackage{amsfonts}

\begin{document}

\begin{frontmatter}

\title{A Note on Gale, Kuhn, and Tucker's Reductions of Zero-Sum Games\tnoteref{label0}}
\tnotetext[label0]{The author thanks Yukihiko Funaki, J\'anos Flesch, Dmitriy Kvasov, Zsombor Z. Meder, and Dan Qin for invaluable comments. She is grateful for the support of Grant-in-Aids for Young Scientists (B) of JSPS No. 17K13707 and Grant for Special Research Project No. 2017K-016 of Waseda University. }

\author{Shuige Liu}
\address{Faculty of Political Science and Economics, Waseda University, 1-6-1
Nishi-Waseda, Shinjuku-Ku, 169-8050, Tokyo, Japan}

\ead{shuige\_liu@aoni.waseda.jp}

\begin{abstract}
Gale, Kuhn and Tucker \cite{gkt50}  introduced two ways to reduce a zero-sum game by packaging some strategies with respect to a probability distribution on them. In terms of value, they gave conditions for a desirable reduction. We show that a probability distribution for a desirable reduction relies on optimal strategies in the original game. Also, we correct an improper example given by them to show that the reverse of a theorem does not hold.
\end{abstract}


\end{frontmatter}


\section{Introduction}
\label{sec:int}
Reducing the size of a game without affecting its instrinsic properties
such as the Nash equilibria, the value, and the optimal strategies, is
desirable from the viewpoint of computation and modelling players' bounded
rationality. In their seminal paper \cite{gkt50}, Gale, Kuhn and Tucker
(GKT) introduced two ways to reduce a zero-sum game by packaging\ some
strategies with respect to some probability distribution on them. A
reduction is desirable if optimal strategies of the original game are
restorable from those of it. The main results there are conditions for a
desirable reduction in terms of value. However, they did not relate that
condition with the probability distribution, which leaves the
practicability of their reductions in question. Also, a $2\times 2$ game intended to
show the reverse of a theorem does not hold is improper, which leaves their
claim unverified.

This note tries to remedy those defects. We show that a probability
distribution for a desirabe reduction relies on optimal strategies of the
original game, which gives a negative answer to the practicability of GKT's reductions.
Also, we show that it is impossible to find a proper example for their claim
in $2\times 2$ games, and give a proper one which is a $4\times 4$ game.
Finally, we discuss some extensions of their results.

\section{Reductions of Zero-Sum Game}
\label{sec:red}

\subsection{Elementary Reduction}

Let $G=(S_{1},S_{2};u_{1},u_{2})$ be a two-person zero-sum game, with
associated mixed strategy sets $M_{1}$ and $M_{2}.$ We denote by $v(G)$
the \emph{value} of $G$, i.e., $v(G)=\max_{x_{1}\in M_{1}} \min_{x_{2}\in M_{2}} u_{1}(x_{1},x_{2})$. $x_{1}\in M_{1}$ is called an \emph{optimal strategy} for
player 1 iff $u_{1}(x_{1},s_{2})\geq v(G)$ for each $s_{2}\in S_{2}.$

GKT introduced two reductions on pure strategy sets. The first one is \emph{%
elementary reduction}. Let $S_{1}^{\prime }\subseteq S_{1}$ with $%
S_{1}^{\prime }\neq \emptyset .$, and $p_{1}$ a probability distribution
over $S_{1}^{\prime }.$ We define $%
G^{p_{1}}=(S_{1}^{p_{1}},S_{2}^{p_{1}};u_{1}^{p_{1}},u_{2}^{p_{1}}),$
where\medskip \newline
\textbf{(i)}. $S_{1}^{p_{1}}=(S_{1}-S_{1}^{\prime })\cup \{\alpha \},$ where $%
\alpha $ is a newly introduced strategy, and $S_{2}^{p_{1}}=S_{2}.$\medskip 
\newline
\textbf{(ii)}. For each $s\in S_{1}^{p_{1}}\times S_{2}^{p_{1}},$%
\begin{equation}
u_{1}^{p_{1}}(s)=\left\{ 
\begin{array}{c}
u_{1}(s) \textrm{ \ \ \ \ \ \ \ \ \ \ \ \ \ \ \ \ \ \ \ \ \ if }s_{1}\neq
\alpha,  \\ 
\Sigma_{s_{1}^{\prime }\in S_{1}^{\prime }}p_{1}(s_{1}^{\prime
})u_{1}(s_{1}^{\prime },s_{2})\textrm{ if }s_{1}=\alpha 
\end{array}%
\right. ,\textrm{ and }u_{2}^{p_{1}}(s)=-u_{1}^{p_{1}}(s).
\end{equation}
We say that $G^{p_{1}}$ is obtained from $G$ by an \emph{elementary player 1
reduction }(\emph{ER1}). The intended meaning of $G^{p_{1}}$ is that we
package\ strategies in $S_{1}^{\prime }$ and introduce a
representative $\alpha $ which is a mechanism that chooses a strategy
randomly in $S_{1}^{\prime }$ by $p_{1}.$
The payoff by choosing $\alpha $
is the expectation on $S_{1}^{\prime }$ by $p_{1}.$ Player 1's mixed strategies in $%
G^{p_{1}}$ are connected to those in $G$ by a mapping $\phi
:M_{1}^{p_{1}}\rightarrow M_{1}$ such that for each $x_{1}^{p_{1}}\in
M_{1}^{p_{1}},$ \medskip \newline
\textbf{(i)}. for each $s_{1}\in S_{1}-S_{1}^{\prime },$ $\phi
(x_{1}^{p_{1}})(s_{1})=x_{1}^{p_{1}}(s_{1});$\medskip \newline
\textbf{(ii)}. for each $s_{1}\in S_{1}^{\prime },$ $\phi
(x_{1}^{p_{1}})(s_{1})=p_{1}(s_{1})x_{1}^{p_{1}}(\alpha ).$\medskip \newline
$\phi $ is injective. It can be seen that for each $%
x_{2}\in M_{2},$ $u_{1}^{p_{1}}(x_{1}^{p_{1}},x_{2})=u_{1}(\phi
(x_{1}^{p_{1}}),x_{2}).$

The purpuse for reduction is to make it easier to find out optimal strategies. Hence we need to know when optimal strategies are preserved, or when optimal strategies of the original game are restorable from those of the reduced game. GKT gave a necessary and
sufficient condition for that as follows:\medskip \newline
\textbf{Theorem \ref{sec:red}.1 (GKT \cite{gkt50})} \textbf{(1)} $%
v(G^{p_{1}})\leq v(G).$\medskip \newline
\textbf{(2)} For each optimal strategies $x_{1}^{p_{1}}$ in $G^{p_{1}}$, $%
\phi (x_{1}^{p_{1}})$ is also an optimal strategy in $G$ if and only if $%
v(G^{p_{1}})=v(G)$.\medskip 

Then it is natural to ask: (1)
For each $S_{1}^{\prime }\subseteq S_{1},$ is there any $p_{1}$ over $S_{1}^{\prime }$ such that $v(G^{p_{1}})=v(G)?$
(2) What property should such a $p_{1}$ satisfy? To answer them, we have the
following statement:$\medskip $\newline
\textbf{Theorem \ref{sec:red}.2.} Let $S_{1}^{\prime }\subseteq S_{1},$ and $%
p_{1}$ be a probability distribution over $S_{1}^{\prime }.$\medskip \newline
\textbf{(1)} If for each optimal strategy $y_{1}$ of $G,$ $\Sigma
_{s_{1}^{\prime }\in S_{1}^{\prime }}y_{1}(s_{1}^{\prime })>0,$ then $%
v(G^{p_{1}})=v(G)$ if and only if there is some optimal strategy $x_{1}$
such that $p_{1}(s_{1})=\frac{x_{1}(s_{1})}{\Sigma _{s_{1}^{\prime }\in
S_{1}^{\prime }}x_{1}(s_{1}^{\prime })}$ for each $s_{1}\in S_{1}^{\prime
}.\smallskip \newline
$\textbf{(2)} If for some optimal strategy $y_{1}$ of $G,$ $\Sigma
_{s_{1}^{\prime }\in S_{1}^{\prime }}y_{1}(s_{1}^{\prime })=0,$ then $%
v(G^{p_{1}})=v(G)$ for any $p_{1}$. 
\footnote{%
The if-part of (1) seems known to GKT. In \cite{gkt50}, p.90, they
said \textquotedblleft [S]uch an elementary reduction
is possible for any decomposition of the matrix $A$, merely by choosing a $P$
depending on some optimal first player strategy for $A.$\textquotedblright
 However, the only-if part of (1) and (2), which are more important, seems
not noticed by them.}\medskip 

To show it, we need the following lemma.\medskip \newline
\textbf{Lemma \ref{sec:red}.1}. Suppose $v(G^{p_{1}})=v(G).$ If for each
optimal strategy $y_{1}$ of $G,$ $\Sigma _{s_{1}^{\prime }\in S_{1}^{\prime
}}y_{1}(s_{1}^{\prime })>0$, then $y_{1}^{p_{1}}(\alpha )>0$ for each
optimal strategy $y_{1}^{p_{1}}$ in $G^{p_{1}}.$\medskip \newline
\textbf{Proof}. Suppose $y_{1}^{p_{1}}(\alpha )=0$ for some optimal strategy 
$y_{1}^{p_{1}}$ in $G^{p_{1}}.$ Since $v(G^{p_{1}})=v(G)$, it follows from
Theorem \ref{sec:red}.1 (2) that $\phi (y_{1}^{p_{1}})$ is an optimal
strategy in $G$, and $\phi (y_{1}^{p_{1}})(s_{1})=y_{1}^{p_{1}}(\alpha
)p_{1}(s_{1})=0$ for each $s_{1}\in S_{1}^{\prime }$.
$\Box$\medskip \newline
\textbf{Proof of Theorem \ref{sec:red}.2:} \textbf{(1)} \textbf{(If)} Let $%
x_{1}$ be an optimal strategy in $G.$ Define $x_{1}^{p_{1}}:S_{1}^{p_{1}}%
\rightarrow \lbrack 0,1]$ by:%
\begin{equation}
x_{1}^{p_{1}}(s_{1}^{p_{1}})=\left\{ 
\begin{array}{c}
x_{1}(s_{1}^{p_{1}})\textrm{ \ \ if }s_{1}^{p_{1}}\in S_{1}-S_{1}^{\prime };
\\ 
\Sigma_{s_{1}^{\prime }\in S_{1}^{\prime }}x_{1}(s_{1}^{\prime })\textrm{ \ \
\ \ if }s_{1}^{p_{1}}=\alpha .%
\end{array}%
\right. 
\end{equation}%
It can be seen that $x_{1}^{p_{1}}\in M_{1}^{p_{1}}$ and $\phi
(x_{1}^{p_{1}})=x_{1}.$ Hence, for each $s_{2}^{p_{1}}\in
S_{2}^{p_{1}}=S_{2},$ $%
u_{1}^{p_{1}}(x_{1}^{p_{1}},s_{2}^{p_{1}})=u_{1}(x_{1},s_{2}^{p_{1}})\geq
v(G)\geq v(G^{p_{1}}).$ Therefore, $x_{1}^{p_{1}}$ is an optimal strategy
for player 1 in $G^{p_{1}},$ and $v(G^{p_{1}})=v(G).$

\textbf{(Only-if)} Suppose that $v(G^{p_{1}})=v(G).$ Let $x_{1}^{p_{1}}$ be
an optimal strategy in $G^{p_{1}}$ and $x_{1}=\phi (x_{1}^{p_{1}}).$ Since
for each optimal strategy $y_{1}$ of $G,$ $\Sigma _{s_{1}^{\prime }\in
S_{1}^{\prime }}y_{1}(s_{1}^{\prime })>0,$ it follows from Lemma \ref%
{sec:red}.1 that $x_{1}^{p_{1}}(\alpha )>0.$ Since for each $s_{1}\in S_{1},$%
\begin{equation*}
x_{1}(s_{1})=\left\{ 
\begin{array}{c}
x_{1}^{p_{1}}(s_{1})\textrm{ \ if }s_{1}\in S_{1}-S_{1}^{\prime } \\ 
p_{1}(s_{1})x_{1}^{p_{1}}(\alpha )\textrm{ \ if }s_{1}\in S_{1}^{\prime }%
\end{array}%
,\right. 
\end{equation*}%
and by Theorem \ref{sec:red}.1 (2), $x_{1}$ is an optimal strategy of $G,$
it can be seen for each $s_{1}\in S_{1}^{\prime },$ $\frac{x_{1}(s_{1})}{%
\Sigma _{s_{1}^{\prime }\in S_{1}^{\prime }}x_{1}(s_{1}^{\prime })}=\frac{%
p_{1}(s_{1})x_{1}^{p_{1}}(\alpha )}{x_{1}^{p_{1}}(\alpha )}=p_{1}(s_{1}).$

\textbf{(2)} Let $p_{1}$ be a probability distribution on $S_{1}^{\prime },$
and $x_{1}$ an optimal strategy of $G$ with $\Sigma _{s_{1}^{\prime }\in
S_{1}^{\prime }}x_{1}(s_{1}^{\prime })=0.$ We define $x_{1}^{p_{1}}\in
M_{1}^{p_{1}}$ by letting $x_{1}^{p_{1}}(\alpha )=0$ and $%
x_{1}^{p_{1}}(s_{1})=x_{1}(s_{1})$ for each $s_{1}\in S_{1}-S_{1}^{\prime }.$
It follows that $x_{1}=\phi (x_{1}^{p_{1}})$ and $v(G^{p_{1}})=v(G).$ $\Box$\medskip 

Theorem \ref{sec:red}.2 casts some doubt on the usefulness of GKT's Theorem %
\ref{sec:red}.1 since it suggests that a desirable reduction needs detailed
and complete information of the optimal strategy in the original game. Since
the purpose of reduction is to make it easier to find an optimal strategy,
why bother to reduce it if we have already known all of them? It may be
helpful in the sense that it allows us to eliminate all irrationalizable
strategies, i.e., iteratedly eliminate strictly dominated strategies (cf.
Pearce \cite{p84}). However, still there are strategies which survive
iterated elimination are assigned 0 in some optimal strategy, and therefore
we cannot exhaust all desirable reductions.

\subsection{Reduction}

GKT extended ER. Let $\mathbb{S}_{i}=\{S_{i1},...,S_{i\ell _{1}}\}$ be a
partition of $S_{i}$ ($i=1,2$). For $t=1,...,\ell
_{1},$ let $p_{1t}$ be a probability distribution over $S_{1t},$ and $%
\mathbf{p}_{1}=(p_{11},...,p_{1\ell _{1}}).$ Define $G^{\mathbf{p}%
_{1}}=(S_{1}^{\mathbf{p}_{1}},S_{2}^{\mathbf{p}_{1}};u_{1}^{\mathbf{p}%
_{1}},u_{2}^{\mathbf{p}_{1}})$ by:$\medskip $\newline
\textbf{(i)} $S_{i}^{\mathbf{p}_{1}}=\{\alpha _{i1},...,\alpha _{i\ell
_{i}}\}$ for $i=1,2;\medskip $\newline
\textbf{(ii)} For $s=(\alpha _{1t_{1}},\alpha _{2t_{2}}),$ $u_{1}^{%
\mathbf{p}_{1}}(s)=\min_{s_{2}\in S_{2t_{2}}}\Sigma _{s_{1}\in
S_{1t_{1}}}p_{1t_{1}}(s_{1})u_{1}(s_{1},s_{2}),$ and $u_{2}^{\mathbf{p}%
_{1}}(s)=-u_{1}^{\mathbf{p}_{1}}(s).\medskip \newline
G^{\mathbf{p}_{1}}$ is said to be obtained from $G$ by a \emph{player 1
reduction }(\emph{R1}). It can be seen that ER1 is a special case of R1,
that is, $\mathbb{S}_{1}=\{S_{1}^{\prime }\}\cup \{\{s_{1}\}\}_{s_{1}\in
S_{1}-S_{1}^{\prime }}$ and $\mathbb{S}_{2}=\{\{s_{2}\}\}_{s_{2}\in S_{2}}$.
Again, player 1's strategies in the two games are connected by a mapping $%
\Phi :M_{1}^{\mathbf{p}_{1}}\rightarrow M_{1}$ where for each $x_{1}^{%
\mathbf{p}_{1}}\in M_{1}^{\mathbf{p}_{1}}$ and $s_{1}\in S_{1t_{1}},\Phi
(x_{1}^{\mathbf{p}_{1}})(s_{1})=p_{1t_{1}}(s_{1})x_{1}^{\mathbf{p}%
_{1}}(\alpha _{1t_{1}}).$

As a parallel to Theorem \ref{sec:red}.1, GKT shown the following
theorem:\medskip \newline
\textbf{Theorem \ref{sec:red}.3 (GKT \cite{gkt50}). (1)} $v(G^{\mathbf{p}%
_{1}})\leq v(G).$\medskip \newline
\textbf{(2)} If $v(G^{\mathbf{p}_{1}})=v(G),$ then for each optimal  
$x_{1}^{\mathbf{p}_{1}}$ in $G^{\mathbf{p}_{1}},$ $\Phi (x_{1}^{\mathbf{p}%
_{1}})$ is also optimal in $G.\medskip $

The difference from Theorem \ref{sec:red}.1 is that in (2), only one
direction holds. GKT argued that the other direction does not hold by the
follwoing example.$\medskip $\newline
\textbf{Example \ref{sec:red}.1 (GKT \cite{gkt50})}. Let $G$ be the Matching
Pennies game, $\mathbb{S}_{1}=\{\{\mathbf{s}_{11}\},\{\mathbf{s}_{12}\}\}$, $%
\mathbb{S}_{2}=\{\{\mathbf{s}_{21},\mathbf{s}_{22}\}\},$ $p_{11}=(1),$ and $%
p_{12}=(1).$ Then we get $G^{\mathbf{p}_{1}}$ as follows:%
\begin{equation}
G=\left[ 
\begin{array}{cc}
1 & -1 \\ 
-1 & 1%
\end{array}%
\right]  \text{ }\underrightarrow{\mathbf{p}_{1}=((1),(1))}\text{ }G^{\mathbf{%
p}_{1}}=\left[ 
\begin{array}{c}
-1 \\ 
-1%
\end{array}%
\right] 
\end{equation}%
Optimal strategy $x_{1}^{\mathbf{p}_{1}}=(\frac{1}{2},\frac{1}{2})$ of $G^{%
\mathbf{p}_{1}}$ maps into optimal strategy $x_{1}=(\frac{1}{2},\frac{1}{2})$
of $G,$ while $v(G^{\mathbf{p}_{1}})=-1<0=v(G).\medskip $

However, this is not a proper example. To show the reverse of Theorem \ref%
{sec:red}.2 (2) does not hold, we need an game where \emph{each} optimal
strategy for player 1 in $G^{\mathbf{p}_{1}}$ maps to an optimal one in $G,$
and $v(G^{\mathbf{p}_{1}})<v(G)$. However, in the above, the former does not
hold. For example, $y_{1}^{\mathbf{p}_{1}}=(0,1)$ is optimal in $G^{\mathbf{p%
}_{1}}$, but $\Phi (y_{1}^{\mathbf{p}_{1}})=(0,1)$ is not optimal in $%
G.\medskip $

Actually, we can show that a proper example cannot be found among $2\times 2$
games, that is, for each $2\times 2$ game, $v(G^{\mathbf{p}_{1}})=v(G)$ if
and only if each optimal strategy $x_{1}^{\mathbf{p}_{1}}$ in $G^{\mathbf{p}%
_{1}}$ corresponds to an optimal one in $G.\footnote{%
A $2\times 2$ game has only four ways to \textquotedblleft
decompose\textquotedblright\ it since for each $S_{i}$, there are only two
ways to partition it, i.e, $\mathbb{S}_{i}=\{S_{i}\},$ $\mathbb{S}%
_{i}=\{\{s_{i1}\},\{s_{i2}\}\}.$ Hence this statement can be shown easily by
checking each case.}$ To find a proper example, we need to go to larger
games. Here we give$\medskip $\newline
\textbf{Example \ref{sec:red}.2}. Consider the following game%
\begin{equation*}
G=\left[ 
\begin{array}{cccc}
3 & 0 & -1 & -1 \\ 
0 & 3 & -1 & 2 \\ 
-1 & -1 & 3 & 0 \\ 
-1 & 2 & 0 & 3%
\end{array}%
\right] .
\end{equation*}%
Let $\mathbb{S}_{i}=\{\{\mathbf{s}_{i1},\mathbf{s}_{i2}\},\{\mathbf{s}_{i3},%
\mathbf{s}_{i4}\}\},$ $i=1,2,$ and $p_{11}=p_{12}=(\frac{2}{3},\frac{1}{3}).$
It can be seen $G^{\mathbf{p}_{1}}$ is just the Matching Pennies game which
has only one optimal  $x_{1}^{\mathbf{p}_{1}}=(\frac{1}{2},\frac{1}{2})$ for
player 1$.$ It can be seen that $\Phi (x_{1}^{\mathbf{p}_{1}})=(\frac{1}{3},%
\frac{1}{6},\frac{1}{3},\frac{1}{6})$ is optimal in $G.$ However, $v(G^{%
\mathbf{p}_{1}})=0<0.5=v(G).\footnote{%
Rather than searching in larger games, here start from the reduced Matching
Pennies game, choose some $\mathbf{p}_{1},$ and construct a game having a
larger value.}\smallskip $

As a parallel to Theorem \ref{sec:red}.2, we have the following statement:$%
\medskip $\newline
\textbf{Theorem \ref{sec:red}.4.} If for each optimal strategies $x_{1}^{%
\mathbf{p}_{1}}$ in $G^{\mathbf{p}_{1}}$, $\Phi (x_{1}^{\mathbf{p}_{1}})$ is
also optimal in $G$, then there exists some optimal strategy $x_{1}$ in $G$
such that for each $S_{1k}\in \mathbb{S}_{1},\medskip $\newline
\textbf{(1)} If $\Sigma _{s_{1k}^{\prime }\in S_{1k}}x_{1}(s_{1k}^{\prime
})>0,$ then $p_{1k}(s_{1k})=\frac{x_{1}(s_{1k})}{\Sigma _{s_{1k}^{\prime
}\in S_{1k}}x_{1}(s_{1k}^{\prime })}$ for each $s_{1k}\in S_{1k};\medskip 
\newline
$\textbf{(2)} If $\Sigma _{s_{1k}^{\prime }\in S_{1k}}x_{1}(s_{1k}^{\prime
})=0,$ then $p_{1k}$ can be any probability distribution over $S_{1k}.$\medskip

Theorem \ref{sec:red}.4 can be proved in a similar way as the only-if part of
Theorem \ref{sec:red}.2 (1) and (2). Its reverse does not hold. Consider the
Matching Pennies game $G$ and $\mathbf{p}_{1}$ in Example \ref{sec:red}.1.
It can be seen that $\mathbf{p}_{1}$ is generated from the optimal strategy $%
(\frac{1}{2},\frac{1}{2})$ of $G.$ However, as shown above, not every
optimal strategy in $G^{\mathbf{p}_{1}}$ corresponds to an optimal one on in 
$G.$

Theorem \ref{sec:red}.4 casts deeper doubt on Theorem \ref{sec:red}.3 than
that by Theorem \ref{sec:red}.2 to Theorem \ref{sec:red}.1 since it
implies that even complete information on optimal strategies in $G$ is not enough; a
desirable R1 requires a selection on optimal strategies and 
\textquotedblleft proper\textquotedblright\ partitions of $S_{1}.$

\section{Some Remarks \label{sec:bis}}

Both reductions can be defined two-sidedly by
iteratedly applying the definitions, and results in this note can be extended
there. Also, they can be extended to zero-sum games with continuum of
strategies. For general $n$-person games, our results hold by replacing
restorability of optimal strategies by that of Nash equilibria, which can be
seen as an extension of Mertens' \cite{m91} small world axiom.

It is wondered whether we can replace the minimum value in the definition of E1 by other ways, e.g., maximum, median, and what is the condition for a desirable reduction. Furthermore, whether there is some desirable reduction which does not rely on the optimal strategies of the original game. We are looking forward for future works in those directions.




\end{document}